\documentclass[10pt,reqno]{amsart}
\usepackage{latexsym,amssymb,amsthm,amsmath}

\theoremstyle{plain}
\newtheorem{theorem}{Theorem}
\newtheorem*{Theorem B}{Theorem B}
\newtheorem*{Theorem A}{Theorem A}

\theoremstyle{remark}

\def\<{\left < }
\def\>{\right >}
\def\({\left ( }
\def\){\right )}

\def\e{\eqref}
\def\p{\partial }

\begin{document}

\thanks\noindent {General Relativity and Gravitation (to appear).}

\vskip.5in

\markboth{B.-Y. Chen}{generalized Robertson-Walker spaces}

\title[Generalized Robertson-Walker spacetimes]{A simple characterization of generalized Robertson-Walker spacetimes}

\author[ B.-Y. Chen]{Bang-Yen Chen}

 \address{Department of Mathematics\\Michigan State University \\619 Red Cedar Road \\East Lansing, MI 48824--1027, USA}

\email{bychen@math.msu.edu}

\begin{abstract} A generalized Robertson-Walker spacetime is the warped product with base an open interval of the real line endowed with the opposite of its metric and  base any Riemannian manifold. The family of generalized Robertson-Walker spacetimes widely extends the one of classical Robertson-Walker spacetimes. In this article we prove a very simple characterization of generalized Robertson-Walker  spacetimes; namely, a Lorentzian manifold is a generalized Robertson-Walker spacetime if and only if it admits a timelike concircular vector field.
\end{abstract}

\keywords{Generalized Robertson-Walker spacetime; Robertson-Walker spacetime; timelike concircular vector field; Lorentzian warped product}

 \subjclass[2000] {Primary 83F05; Secondary 53C25}

\maketitle

\section{Introduction}

An $n$-dimensional generalized Robertson-Walker (GRW) spacetime with $n\geq 3$ is  a Lorentzian manifold which is a warped product manifold $I\times_f F$ of an open interval $I$ of the real line ${\mathbb R}$ and a Riemannian $(n-1)$-manifold $(F,g_F)$  endowed with the Lorentzian metric
\begin{align}\label{1}\bar g=-\pi^*_I (dt^2)+f(t)^2 \pi^*_F(g_F),\end{align}
where $\pi_I$ and $\pi_F$ denote the projections onto $I$ and $F$, respectively, and $f$ is a positive smooth function on $I$. In a classical Robertson-Walker (RW) spacetime, the fiber is three dimensional and of constant sectional curvature, and the warping function $f$ is arbitrary. 

A Robertson-Walker spacetime obeys the cosmological principle, i.e.,  it is spatially homogeneous and spatially isotropic. As said before,  GRW spacetimes widely extend RW spacetimes and include, among others, the Lorentz-Minkowski spacetime, the Einstein-de Sitter spacetime, the Friedmann cosmological models, the static Einstein
spacetime and the de Sitter spacetime. GRW spacetimes obey the well-known Weyl hypothesis, i.e.,  the world lines should be everywhere orthogonal to a family of spacelike slices. 

Contrary to Robertson-Walker spacetimes, GRW spacetimes are not necessarily spatially-homogeneous. Spatially-homogeneity is reasonable as a first approximation of the large scale structure of the universe, but it could not be appropriate when one considers a more accurate scale. On the other hand, small deformations of the metric on the fiber of classical RW spacetimes fit into the class of  GRW spacetimes. Further, GRW spacetimes appear as a privileged class of inhomogeneous spacetimes admitting an isotropic radiation (cf. e.g. \cite{Sa98}).

GRW spacetimes have been investigated in \cite{Ale,Ali,Ali2,CRR,DK,RRS1,RRS2,Sa98,Sa} among others. In particular, a  global characterization of GRW spacetimes in term of a timelike and spatially conformal vector field satisfying certain natural conditions was obtained by M. S\'anchez in \cite{Sa98}. Also, several  characterizations of GRW spacetimes in term of timelike gradient conformal vector fields were established by M. Caballero, A. Romero and R. M.  Rubio in \cite{CRR}.

A. Fialkow introduced in \cite{F} the notion of {\it concircular vector field} on a Riemannian manifold $M$ as a vector field $v$ which satisfies
\begin{align}\label{2} \nabla_X v=\mu X\end{align}
for  vectors $X$ tangent to $M$, where $\nabla$ denotes the Levi-Civita connection of $M$ and $\mu$ is a function on $M$. Observe that a vector field $v$ satisfying \e{2} is conformal with $\mathcal L_v g=2\mu g$, where $g$ is the metric tensor. Moreover, if $\omega$ denotes the 1-form metrically equivalent to $v$,  then
$d\omega=0$.  In particular, if $v$ has no zero, in the Riemannian case, or it is
timelike in the Lorentzian case, the distribution $v^\perp$ is always integrable.

Concircular vector fields are also known as geodesic fields  in literature since   unit speed integral curves of such vector fields are geodesics. Concircular vector fields appeared in the study of concircular mappings, i.e., conformal mappings
preserving geodesic circles \cite{Y}.  Concircular vector fields play an important role in the  theory of projective and conformal transformations.  Such vector fields have interesting applications in general relativity, e.g.,  trajectories of timelike concircular fields in the de Sitter model determine the world lines of receding or colliding galaxies satisfying the Weyl hypothesis  \cite{T}.

The purpose of this article is to establish the following very simple characterization of generalized Robertson-Walker spacetimes in term of timelike concircular vector field.

\begin{theorem} A Lorentzian $n$-manifold with $n\geq 3$ is a generalized Robertson-Walker spacetime if and only if it admits a timelike concircular vector field.
\end{theorem}

\section{Proof of the theorem 1}

For general references on pseudo-Riemannian submanifolds, we refer to \cite{book,book14,O}.
Assume that $M$ is a Lorentzian $n$-manifold with $n\geq 3$.  Suppose that $M$ admits a timelike concircular vector field. 
Let us put
\begin{align} \label{3} v=\varphi e_1,\end{align}
where $e_1$ is a unit  timelike vector field in the direction of $v$. 
Let us extend $e_1$ to an orthonormal frame $e_1,e_2,\ldots,e_{n}$ on $M$ so that $e_2,\ldots,e_{n}$ are orthonormal spacelike vector fields on $M$. Define the connection forms $\omega_i^j\, (i,j=1,\ldots,n)$  by
\begin{align} \label{4}& \nabla_X e_i=\sum_{j=1}^n \epsilon_j\omega_i^j(X)e_j,\;\;i=1,\ldots,n,\end{align} where $\epsilon_1=-1$ and $\epsilon_2=\cdots=\epsilon_n=1$.
From Cartan's structure equations, we have
$d\omega^i=-\sum_{j=1}^n \epsilon_j \omega^i_j\wedge \omega^j$ for $ i=1,\ldots,n$.

It follows from \e{2} and a direct computation that the curvature tensor $R$ of $M$ satisfies
\begin{equation}\begin{aligned} \label{6}  R(e_i,v)v&=\nabla_{e_i}\nabla_v v-\nabla_v\nabla_{e_i} v-\nabla_{[e_i,v]}v \\&=(e_i\mu)v-(v\mu)e_i,\;\; \end{aligned}\end{equation} 
for $i=2,\ldots,n$, where $\mu$ is defined by \e{2}. From \e{6} we get 
 \begin{align}\label{7}e_2\mu=\cdots=e_n\mu=0.\end{align}  
 Thus the gradient $\nabla \mu$ of $\mu$ is a vector field parallel to $v$.
From \e{2} with $X=e_1$ and \e{3} we find 
$$\mu e_1=\nabla_{e_1}(\varphi e_1)=(e_1\varphi)e_1+\varphi \nabla_{e_1}e_1,$$
which gives
 \begin{align}\label{8}&e_1\varphi=\mu,\;\; \\&\label{9}\nabla_{e_1}e_1=0.\end{align}  It follows from  \e{9}  that the integral curves of $e_1$ are geodesics in $M$. Therefore the distribution $\mathcal D_1={\rm Span}\{e_1\}$ is a totally geodesic foliation, i.e., $\mathcal D_1$ is an integrable distribution whose leaves are totally geodesic one-dimensional negative definite submanifolds of $M$.
Let us define another distribution by putting $\mathcal D_2={\rm Span}\{e_2,\ldots,e_n\}$, i.e., $\mathcal D_2=v^\perp$. 

From \e{2}  with $X=e_i$ $(i=2,\ldots,n)$ and  \e{3}, we have
$$\mu e_i=\nabla_{e_i}(\varphi e_1)=(e_i\varphi)e_1+\varphi \nabla_{e_i}e_1,$$
which implies that \begin{align}\label{10}&e_2\varphi=\cdots=e_n\varphi=0,\;\;\\&\label{11} \varphi \nabla_{e_i}e_1=\mu e_i.\end{align}
From \e{4} and  \e{11} we obtain
\begin{align}\label{12}&\omega_i^1(e_j)=\frac{\mu}{\varphi}\delta_{ij},\;\; 2\leq i,j\leq n,\end{align}
where $\delta_{ij}$ stands for the Kronecker delta function.

It follows from \e{12} that $\mathcal D_2$ is an integrable distribution whose leaves are totally umbilical in $M$. Moreover, the mean curvature of the leaves of $\mathcal D_2$ is given by $\mu/\varphi$. Since leaves of $\mathcal D_2$ are spacelike hypersurfaces, it follows from \e{7} and \e{10} that the mean curvature vector fields of leaves of $\mathcal D_2$ are parallel in the normal bundle in $M$. Thus $\mathcal D_2$ is a spherical foliation, i.e., $\mathcal D_2$ is an integrable distribution whose leaves are totally umbilical spacelike hypersurfaces with constant mean curvature. Consequently, by a result of S. Hiepko \cite{H} (or Theorem 4.4 of \cite[page 90]{book}) (see also \cite[Corollary 1]{PR}), we conclude that $M$ is an open portion of a warped product $I\times_{f}  F$, where $f$ is a function on $I$, $\p/\p t=e_1$, and $F$ is a Riemannian $(n-1)$-manifold. Therefore the sectional curvature of $M$ satisfies
\begin{align}\label{13} K(e_1,e_i)=\frac{f''(t)}{f(t)}\end{align}
for $i=2,\ldots,n$.

On the other hand, it follows from \e{3} and \e{6} that 
\begin{align}\label{14} \varphi^2 K(e_1,e_i)=v(\mu)=\varphi \mu'(t)\end{align}
for $i=2,\ldots,n$. Thus, after combining \e{14} with \e{8} and \e{13}, we obtain
\begin{align}\notag \frac{f''(t)}{f(t)}=\frac{\mu'(t)}{\varphi}=\frac{\varphi''(t)}{\varphi(t)}.\end{align}
Consequently, if we choose $f(t)=\varphi(t)$, then $M$ is an open portion of the Lorentzian warped product manifold $I\times_{f}  F$ with $f(t)=\varphi(t)$.

Conversely, let us consider a Lorentzian manifold which is a warped product manifold of the form:  
$I\times_{f} F$,
where $(F,g_F)$ is a Riemannian manifold so that the metric tensor $\bar g$ of $I\times_{f} F$ is given by \e{1} with $f$ being a positive function on $I$.
 Consider the timelike vector field given by
$v= f(t)\frac{\p}{\p t}.$
Then it follows from Proposition 7.35(2) of \cite[page 206]{O} or Proposition 4.1 of \cite[page 79]{book} and a direct computation that  $v$ satisfies condition \e{2} with  $\mu=f'(t)$. Therefore $v$ is a timelike concircular vector field.
This completes the proof of Theorem 1.

\vskip.1in 
\noindent {\bf Acknowledgements.} The author thanks the referees for their useful suggestions for improving the presentation of this paper.

\end{document}